\documentclass[%
 reprint,
superscriptaddress,
showpacs,
 amsmath,amssymb,
 aps,
]{revtex4-2}

\usepackage{rotating}
\usepackage{stackengine}
\usepackage{graphicx}
\usepackage{color}
\usepackage{verbatim}
\usepackage{tikz}
\usepackage{tkz-graph}
\usepackage{braket}
\usepackage{subfigure}
\definecolor{Blue}{rgb}{0,0,1}
\definecolor{Red}{rgb}{1,0,0}
\definecolor{Black}{rgb}{0,0,0}
\definecolor{Green}{rgb}{0.0, 0.5, 0.0}

\begin{document}

\preprint{APS/123-QED}

\title{Perfect state transfer in quantum photonic networks based on Fourier modes} 
\author{Sunita Meena} 
\email{sunita40$_$sps@jnu.ac.in}
\affiliation{School of Physical Sciences, Jawaharlal Nehru University, New Delhi 110067, India}
\author{Sugar Singh Meena} 
\email{sugar15$_$sps@jnu.ac.in}
\affiliation{School of Physical Sciences, Jawaharlal Nehru University, New Delhi 110067, India}
\author{Paulo A Brandão} 
\email{paulo.brandao@fis.ufal.br}
\affiliation{Instituto de F\'isica, Universidade Federal de Alagoas, Macei\'o, Brasil}
\author{Amit Rai} 
\email{amitrai@mail.jnu.ac.in }
\affiliation{School of Physical Sciences, Jawaharlal Nehru University, New Delhi 110067, India}

\begin{abstract}
We propose a quantum network consisting of optical waveguides in the linear regime for quantum state transfer. The circular topology of our network introduces novel functionalities that enable us to analytically identify the conditions under which perfect state transfer (PST) is achievable. We utilize the properties of the Fourier modes, in particular the zero Fourier modes, which provide a protected subspace for the efficient propagation of quantum states, resulting in PST to the diametrically opposite site in a network with any number of sites $N = 4n$. The coupling profiles in the photonic network modulate the number of zero-energy eigenmodes, with uniform couplings yielding more than evanescent ones, confirming that the expedition of observed PST originates from the collapse of the eigenvalue spectrum into three distinct eigenvalue blocks, including $N/2$ manifolds of zero Fourier modes. We investigate PST in both discrete- and continuous-variable input regimes, using single photon state, Schr\"odinger cat states,  and two-mode squeezed vacuum state. Our findings apply to the engineering of quantum networks and photonic lattices, paving the way for applications in controlled routing in integrated quantum circuits.
\end{abstract}

\maketitle

\section{Introduction}
\label{intro}
Information transmission through quantum channels from one end to the other, known as quantum state transfer (QST) \cite{bose2003}, is essential for quantum networks and circuits with applications in quantum routing \cite{kristjansson2024quantum, abane2025entanglement, hahto2025transfer}, long-distance secure quantum communication \cite{fukui2021all, chen2021integrated}, quantum sensor networking \cite{eldredge2018optimal} and distributed quantum computing \cite{caleffi2024distributed, main2025distributed}. 
Robust and high-fidelity QST in large-scale physical systems is essential to advance these quantum technologies. From the foundational work on spin chains \cite{bose2003, christandl2004perfect}, the idea of quantum state transfer attracted significant interest and has been implemented in various physical systems, such as photonic lattices \cite{perez2013coherent, himmel2026state}, superconducting circuits \cite{li2018perfect, xiang2024enhanced}, and quantum walks on graphs \cite{vstefavnak2017perfect}. However, maintaining efficient, high-fidelity QST in complex quantum photonic networks and integrating them onto a single chip remains an open challenge.

In the discrete-variable (DV) regime, quantum state distribution using single photons has seen rapid progress, with numerous protocols developed \cite{shomroni2014all, saha2023routing}. Optimal schemes for distributing entangled states have also been proposed \cite{dai2020optimal, halder2024optimal} and demonstrated \cite{thomas2025high}. The controllable distribution of quantum states in the continuous-variable (CV) regime has been demonstrated \cite{wang2021controllable}. Hybrid quantum systems for QST between CV and DV have also been explored \cite{hastrup2022universal, liu2026hybrid}. Waveguide lattices provide a robust, scalable, and low-loss platform for a range of studies and analyses, including simulations of condensed matter physics effects, state transfer, and optical quantum information processing \cite{szameit2010discrete, yang2025programmable}. Arrays of optical waveguides with evanescent nearest-neighbor couplings have been used to transfer quantum states between different sites \cite{perez2013coherent, perez2013perfect, chapman2016experimental}. Although previous work has addressed planar waveguide arrays, other topologies could offer novel functionalities. In particular, circular geometry of waveguide arrays is pivotal for applications such as optical switching, power distribution, optical state transfer, and robust multipartite entanglement generation \cite{longhi2007light, hudgings2002design, rai2022transfer, meena2026robust}. However, studies of QST in circular arrays \cite{rai2022transfer, beder2024quantum} rely primarily on numerical simulations with a small number of waveguides. As the number of modes increases, achieving perfect state or high-fidelity transfer in these systems requires significantly longer propagation distances, which are prone to optical losses and hard to fabricate \cite{Paschotta_2019_propagation_losses}. Engineering circular arrays for a realistic quantum photonic network with any number of waveguides $N$ to enhance the information capacity of communication channels using PST is still missing.

In this work, we address this question and propose an $N$-waveguide quantum photonic network with uniform long-range couplings that exhibit exact PST under physically realizable conditions. Our approach produces analytical solutions using the discrete symmetries of Fourier modes for the PST, providing physical insights into the dynamics of the state in the $N$-waveguide network. We show analytically that when the interaction range ($R$) satisfies $R=N/2-1$, and the system size obeys $N=4n$ $(n = 1, 2,...)$, the eigenvalue spectrum collapses into three distinct blocks. This spectral collapse produces the necessary phase across all eigenmodes to attain perfect transfer of the input quantum states. Interestingly, our waveguide network exhibits a unique feature: it contains $N/2$ manifolds of supermodes or Fourier modes with zero eigenvalues. The eigenmodes with zero eigenvalue, zero Fourier modes, provide a protected subspace for the efficient propagation of optical quantum states and play a significant role in PST to the antipodal modes of the network. The number of zero Fourier modes depends on the network's coupling profile: uniform long-range couplings yield more than evanescent ones. We demonstrate that our waveguide network with uniform long-range couplings for PST is independent of the number of waveguides $N$ and outperforms schemes with nearest-neighbor and evanescent long-range couplings, for which PST efficiency degrades with increasing system size. We also observed that uniform long-range interactions expedite PST, whereas long-range evanescent interactions achieve near-perfect transfer only at significantly longer propagation distances. Beyond the transfer of single excitation in the DV regime, we extend our analysis to CV quantum-state transfer. In particular, we examine the perfect transfer of the injected single-photon state, two-mode squeezed vacuum (TMSV) state, and Schr\"odinger cat states. We establish a unified framework for DV and CV quantum state transfer in quantum photonic networks, highlighting the role of Fourier modes in routing both excitations and quantum correlations. Finally, we discuss experimental feasibility and show that auxiliary-mediated interactions can effectively generate uniform long-range couplings, providing a realistic pathway for implementing the proposed scheme in integrated photonic platforms and related quantum architectures \cite{tillmann2013experimental, carolan2015universal, wang2020integrated}.

The paper is organized as follows. In Sec.\,\ref{theory}, we introduce the quantum photonic network and discuss the evolution of its Fourier modes. In Sec.\,\ref{results}, we identify the conditions to get analytical solutions for PST, present numerical results of single- and multimode quantum states, and analyze the effects of evanescent coupling on PST. We discuss the feasibility of our proposed quantum network using auxiliary-mediated coupling schemes in Section~\ref{feasibility}. Finally, Sec.\,\ref{discussionAndCon} summarizes the results and outlines future extensions of our work.


\section{Theory}
\label{theory}
In this section, we develop the theoretical framework for the proposed quantum photonic network by introducing the Hamiltonian with long-range couplings and diagonalizing it using the discrete Fourier matrix. Here, we show how long-range couplings can improve faithful communication by transferring quantum states perfectly from one end to the other in a quantum network. Then we discuss the linear supermodes or Fourier modes of the network and the collapse of their eigenvalue spectrum, which plays a significant role in PST. Moreover, we provide analytical solutions for the PST in the network using the symmetries of the Fourier modes. We examine two coupling regimes for state transfer: the first involves uniform long-range couplings, whereas the other involves evanescent long-range couplings modeled by exponentially decaying interaction strengths. We study the dynamics of discrete- and continuous-variable states in the photonic networks and present numerical results for perfect transfer. Throughout this work, we have set $\hbar=1$.

\subsection{$N$-waveguide network model}
We consider a $N$-waveguide network consisting of identical femtosecond-laser-written waveguides arranged in a circular (ring) geometry. Any two waveguides in the network are assumed to have coupling, although some coupling strengths may later be set to zero. Each waveguide is modeled as a Bosonic mode with annihilation (creation) operator \(\hat {a}_{j} \) (\(\hat {a}_ {j}^\dagger \)), obeying the commutation relations
\([\hat a_j,\hat a_k^\dagger]=\delta_{jk}\). The tight-binding Hamiltonian can describe the N-waveguide network with long-range interactions

\begin{equation}
H = \sum_{r=1}^{R} C_r 
\sum_{j=0}^{N-1}
\left(
\hat a_j^\dagger \hat a_{j+r}
+ \hat a_{j+r}^\dagger \hat a_j
\right),
\label{eq:H_general_simple}
\end{equation}
where \(C_r\) denotes the coupling strength between the waveguides separated by \(r\) neighbors along the ring. The following Heisenberg equation is obtained
\begin{equation}
\frac{d \hat{a}_{j}}{dz}
= -i
\sum_{r=1}^{R} C_r
\left(
\hat a_{j+r} + \hat a_{j-r}
\right) \label{eq:HEOM}
\end{equation}
where $N+j \equiv j~ (mod ~N)$, i.e. $ \hat{a}_{N} \equiv \hat{a_{0}}$, $\hat{a_{N+1}} \equiv \hat{a_{1} }$, and $j = 0,1,..., N-1$ is the individual mode index. The above propagation equation describes network dynamics and can be solved numerically for a given set of coupling parameters or analytically for a waveguide network with small $N$. However, increasing the number of waveguides $N$ in the network makes it more complex, and it is difficult to gain physical insight from numerical or small-scale analytical solutions. The problem of propagation (\ref{eq:HEOM}) can be simplified using the eigenmodes of the proposed photonic network -- the linear or propagation supermodes. Interestingly, for our network, these supermodes are discrete Fourier modes \cite{hudgings2002design}. In the next sections, we introduce some properties of these discrete Fourier modes and show how they are useful for finding the analytical solutions for PST in the quantum photonic network.

\subsection{Fourier modes of the network}
For a waveguide network with uniform all-to-all couplings, the linear coupling matrix of the system becomes highly symmetric. It can be represented by an $N \times N$ circulant matrix \text{C} with vanishing diagonal elements and all off-diagonal elements being equal to $C$, as $C_{i,j} = C(1-\delta_{i,j})$ (where $\delta$ is the Kronecker delta
function). Due to their symmetry and translation invariance, network geometries with a circulant matrix C, allow us to obtain analytical solutions for periodic systems. This matrix \text{C} can be diagonalized using the discrete Fourier matrix $\mathcal{S}$; a complex symmetric and unitary matrix with elements $\mathcal{S}_{j,p}$. The Fourier and individual mode bases are related by  
\begin{align}
\hat b_p = \mathcal{S}_{j,p}\,\hat a_j, \qquad p = 0,1,..., N-1
\label{eq:supermodes}
\end{align}
\text{with}
\begin{equation*}
   \mathcal{S}_{j,p} = \frac{1}{\sqrt{N}}\sum_{j=0}^{N-1}
e^{\frac{i2\pi jp}{N}} 
\end{equation*}
In the Fourier mode basis, the Hamiltonian is simplified to
\begin{equation}
\hat H = \sum_{p=0}^{N-1} \lambda_p\, \hat b_p^\dagger \hat b_p 
\end{equation}
with eigenvalue spectrum
\begin{equation}
\lambda_p = 2 \sum_{r=1}^{R} C_r
\cos\!\left(\frac{2\pi p r}{N}\right)
\label{eq:dispersion_general}
\end{equation}
The evolution of the Fourier mode operators $\hat {b}_{p} $ in the Heisenberg picture is given by
\begin{equation}
\hat b_p(z) = e^{-i\lambda_p z}\,\hat b_p(0)
\end{equation}
and the transition probability amplitude between waveguides $j$ and $l$ is governed by
\begin{equation}
U_{jl}(z) = \frac{1}{N} \sum_{p=0}^{N-1}
e^{-i\lambda_p z}
e^{i\frac{2\pi p}{N}(j-l)} 
\label{eq:U_general}
\end{equation}


\section{Results and Analysis}
\label{results}

\subsection{Analytical results for PST}
\label{ANALYTICALPST}
We first derive the analytical condition for PST in the waveguide quantum network with long-range uniform coupling. We consider a photonic network consisting of $N$ waveguides with uniform coupling $C_r=C$ and interaction range $R=N/2-1$. The eigenvalues of the Hamiltonian are
\begin{equation}
\lambda_p = 2C \sum_{r=1}^{N/2-1} 
\cos\!\left(\frac{2\pi p r}{N}\right)
\end{equation}
Using the geometric identity of discrete Fourier sums (see Appendix A), this simplifies to

\begin{equation}
\lambda_p = -C\left[1 + (-1)^p \right]
\label{eqn:LambdaPGen}
\end{equation}
with an additional eigenvalue $\lambda_0 = C(N-2)$. The eigenvalue spectrum, therefore, collapses into three distinct energy blocks

\begin{align}
\lambda_p &= 0 \quad (p \ \text{odd}) \\
\lambda_p &= -2C \quad (p \ \text{even},\, p\neq0) \\
\lambda_0 &= C(N-2)
\label{eqn:Lambda specific p values}
\end{align}
and the supermode spectrum becomes
\begin{equation}
\{C(N-2), \; 0, \; -2C\}
\end{equation}

This spectral collapse produces $N/2$ degenerate zero Fourier modes, which play a crucial role in the dynamics, as they remain protected during propagation. Expanding the propagator in Eq.~\ref{eq:U_general} in terms of the collapsed eigenvalue spectrum,

\begin{align}
U_{jl}(z)
&=
\frac{1}{N} \Bigg[
e^{-iC(N-2)z}+\sum_{p\,\text{odd}} e^{i\frac{2\pi p}{N}(j-l)}
\nonumber\\
&\quad \qquad \qquad \qquad +
e^{i2Cz}
\sum_{\substack{p\,\text{even}\\ p\neq 0}} e^{i\frac{2\pi p}{N}(j-l)}
\Bigg]
\end{align}
Let $d = j-l$. The discrete Fourier sums over odd and even indices can be evaluated analytically. The propagator simplifies to
\begin{align}
U_{jl}(z)
&=
\frac{1}{N}
\Bigg[
e^{-iC(N-2)z}
+
\frac{N}{2}\big(\delta_{d,0}-\delta_{d,\frac{N}{2}}\big)
\nonumber\\
&\qquad \qquad 
+
e^{i2Cz}
\left(
\frac{N}{2}\big(\delta_{d,0}+\delta_{d,\frac{N}{2}}\big)-1
\right)
\Bigg]
\label{eq:U_simplified}
\end{align}
This expression shows that the dynamics is dominated by two special points: the initial site ($d=0$) and the antipodal site ($d=N/2$). For lattices with $N=4n$, the antipodal point is uniquely defined and plays a central role in transport. For $d = N/2$, Eq.~(\ref{eq:U_simplified}) reduces to
\begin{equation}
U_{j,l+\frac{N}{2}}(z)
=
\frac{1}{N}
\left[
e^{-iC(N-2)z}
-
\frac{N}{2}
+
e^{i2Cz}\left(\frac{N}{2}-1\right)
\right]
\end{equation}
PST occurs when $|U_{j,l+\frac{N}{2}}(z)|=1$. This is achieved when all spectral components contribute in such a way that phase factors synchronize for the required PST. For $N=4n$, this condition is satisfied at
\begin{equation}
Z_{\text{PST}} = \frac{(2s+1)\pi}{2C}, \quad (s = 0,1,2,...)
\end{equation}
for which
\begin{equation}
e^{i2CZ_{PST}} = -1 
\end{equation}
\begin{equation}
e^{-iC(N-2)Z_{PST}} = (-1)^{\frac{N}{2}-1}=-1
\end{equation}
Substituting these into the propagator yields
\begin{equation}
\langle j|U(Z_{\text{PST}})|l\rangle
= -\delta_{j,l+N/2}
\label{PSTNegativePhase}
\end{equation}
which implies
\begin{equation}
|U_{j,l+\frac{N}{2}}(Z_{\text{PST}})|^2 = 1
\end{equation}

Thus, for $N=4n$, the system exhibits perfect state transfer to the antipodal site at a propagation distance $Z_{PST}$. This result shows that PST in our proposed network arises from a highly structured supermode spectrum comprising three energy blocks. The global spectrum is therefore more important for observing PST rather than local coupling strength and coupling range, thereby establishing a precise analytical condition for long-range quantum state transfer.


\subsection{Numerical verification of PST}

We now numerically verify the analytical spectral-collapse condition derived in Sec.\,\ref{ANALYTICALPST}. As established, a highly structured eigenvalue spectrum enables the necessary phases to be attained for PST.
\subsubsection{Single Photon PST}
\label{SPPST}
\begin{figure*}[htbp]
\centering
\includegraphics[width=0.45\linewidth]
{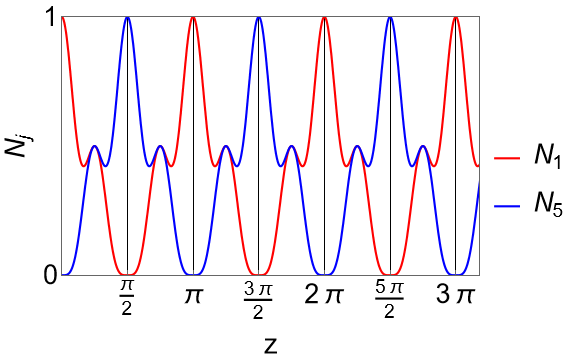}
\put(-200,127){(a)}
\hfill
\includegraphics[width=0.45\linewidth]{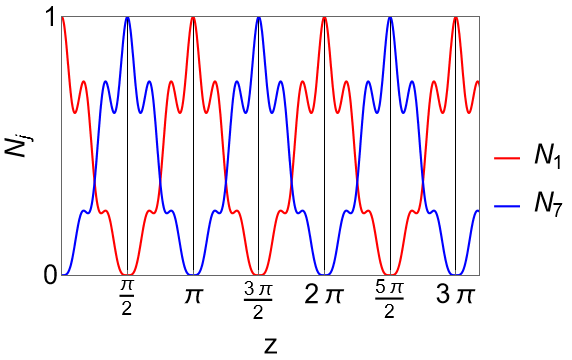}
\put(-200,127){(b)}
\caption{Single photon state propagation in the waveguide quantum network. (a) To verify perfect transfer, we investigate the photon number for the network with $N=8$ and $R=3$. We obtained PST from input mode $1$ to the diametrically opposite mode $5$, i.e., from $1\rightarrow5$. (b) For $N=12$ and $R=5$, we obtained PST from mode $1\rightarrow7$.}
\label{fig:PST_all}
\end{figure*}

A single photon state is the most non-classical of quantum states. The single-photon fidelity directly influences the ability to transmit secure quantum bits over a predefined distance. It can be represented as a single excitation 
\begin{equation}
  \ket{\psi_{SP}}= \hat{a}^\dagger_m\ket{0}
\end{equation}
where $m$ is the excitation mode of the quantum network. Since it has only one photon, fidelity and photon-number behavior are equivalent. We define the photon number in mode $j$ as $N_j=\braket{\hat{a}^{\dagger}_j\,\hat{a}_j}$. From numerical simulations of photon number in a single photon state, Fig. \ref{fig:PST_all} shows the transport dynamics for the waveguide quantum network with $N=8$ and $N=12$. All cases correspond to $C=1$, with the input supplied in the $l=1$ mode, and we observe the state transfer to the antipodal mode of the network.  In all cases, perfect transfer occurs exclusively when the interaction range is set to $R=N/2-1$, i.e., the regime where the eigenvalue spectrum collapses into three distinct eigenvalue sets. For $N=8$, PST occurs at $R=3$ with transfer of state from mode $1\rightarrow5$. For $N=12$, PST is observed at $R=5$ with transfer of state in from mode $1\rightarrow7$. Away from these specific coupling ranges, the spectral structure is lifted, leading to rapid degradation of transfer fidelity. This confirms that PST is governed by global spectral interference rather than local coupling strength and coupling range alone.

Single photon states have important potential applications, such as linear optical quantum computing (LOQC), quantum key distribution (QKD), and quantum metrology, which motivate the study of single photons \cite{scheel2009single}. Moreover, PST of single-photon states has applications in optical quantum information processing, typically involving single photons in single spatio-temporal modes, and requires that the fidelity be close to unity \cite{oxborrow2005single}. 

\subsubsection{Schrodinger Cat States PST}
\label{SCSPST}
Quantum states with strong nonclassical properties, prepared in superpositions of classically distinguishable states, are called Schrodinger cat states (SCS) \cite{gerry1997quantum}. These states are a superposition of two coherent states of equal amplitude but have a phase of $\pi$ \cite{gerry1997quantum,gerry2023introductory,monroe1996schrodinger}. The general form of SCS is 
\begin{equation}
  \ket{\psi_{SCS}}= \mathcal{N}( \ket{\alpha}+ e^{i \phi}\ket{-\alpha})
\end{equation}
where the normalization factor has the form
\begin{equation}
  \mathcal{N}=( 2+2\,  e^{-2\alpha^2}\, \cos{\phi} )^{-1/2}.
\end{equation}
For simplicity, we assume $\alpha$ to be real and the relative phase $\phi$ can take any arbitrary values. Three important cat states corresponding to $\phi=0,\pi/2$, and $\pi$ are named as even coherent states, Yurke–Stoler (YS) states, and odd coherent states, respectively \cite{gerry2023introductory,yurke1988dynamic}. Although these cat states have been generated and studied in some photonic lattices \cite{lewenstein2021generation,urbieta2025dynamics,urbieta2026robustness}, PST with unit fidelity in networks with larger system size $N$, more importantly, in waveguide-based circular quantum photonic networks remains unexplored. Using the analytical Fourier-mode framework, we have explored the PSTs of these states.

Since SCS are not single photon states, a complete transfer of photons does not indicate a transfer of the state itself. To trace the PST in our photonic network, we define a standard quantity, named fidelity. Fidelity is a mathematical parameter that quantifies the degree of exactness between two quantum states, defined as $F=|\braket{\psi{|\psi(z)}}{}|^2$. Upon solving further, the analytical expression, as reported in \cite{urbieta2025dynamics}, is-

\begin{equation}
    F=|2\,\mathcal{N}^2 e^{-|\alpha|^2} \Big[ e^{ e^{i \beta z} \, U^{*}_{j,l}(z)   \,|\alpha|^2}+ \, \cos{\phi} \,e^{ -e^{i \beta z }\, U^{*}_{j,l}(z)   \,|\alpha|^2}\Big]|^2
    \label{FidCat}
\end{equation}

where $\beta$ is the free propagation constant and is considered zero as it contributes a global phase only. Eq. (\ref{FidCat})  is explicitly dependent on the relative phase $\phi$ and $|\alpha|^2$ associated with $\alpha$ values. This phase dependency would affect the PST of three kinds of SCS. 

\begin{figure*}
\centering
\begin{subfigure}
    \centering
    \includegraphics[width=3.4in, height=1.9in]{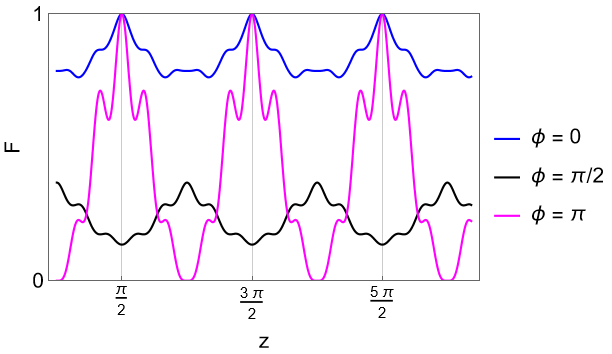}
    \put(-222,122){(a)}
\end{subfigure}
\hfill
\begin{subfigure}
    \centering
    \includegraphics[width=3.4in, height=1.9in]{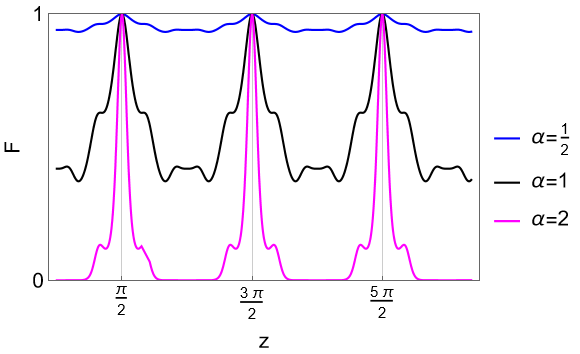}
    \put(-222,122){(b)}
\end{subfigure}
\vspace{0.5cm}
\begin{subfigure}
    \centering
\includegraphics[width=3.4in, height=1.9in]{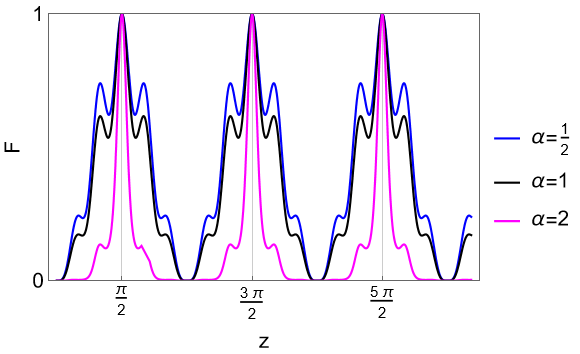}
\put(-222,122){(c)}
\end{subfigure}
\hfill
\begin{subfigure}
    \centering
\includegraphics[width=3.4in, height=1.9in]{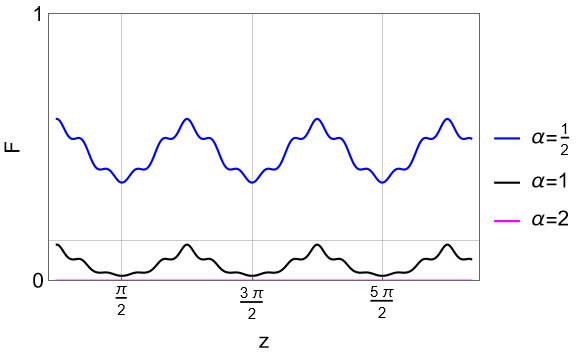}
\put(-222,122){(d)}
\end{subfigure}
\caption{For a waveguide quantum network with $N=12$ and $R=5$, we have plotted the fidelity vs propagation distance for SCS. We obtained perfect transfer from input mode $1$ to the antipodal or diametrically opposite mode $7$, i.e., from $1\rightarrow7$. (a) For SCS with $\alpha=1/\sqrt{2}$ and three different values of $\phi=0,\pi/2, \pi$. We also examine (b) Even cat state with $\phi=0$, (c) Odd cat state with $\phi=\pi$, and (d) YS cat state with $\phi=\pi/2$ for three different values of $\alpha=1/2, 1$, and $2$.}
\label{fig:ResultCatState}
\end{figure*}

From our numerical simulations in Fig. (\ref{fig:ResultCatState}), it is evident that PST takes place for odd and even cat states, although it reaches a maximum of $\approx0.36$ values for YS cat states. The PST of SCS in our quantum photonic network highlights the sensitivity of the Fourier mode framework for the YS state and its robustness for odd and even cat states. From these plots, we deduce the optimal $\alpha$ values and phase values. To achieve a higher overall average fidelity, one should consider smaller values of $\alpha$ and $\phi=0$ \cite{urbieta2025dynamics,longhi2013quantum}. While larger values of $\alpha$ and $\phi=\pi/2$ have to be avoided. For YS cat states, a smaller $\alpha$ ($=1/2$) value makes the fidelity reach a maximum of $\approx0.6$. This is further discussed in Appendix D (also see Fig.~\ref{fig:catStateFgeneral}).

PST of Schr\"odinger cat states enables high-precision quantum metrology ~\cite{joo2011quantum}, essential for quantum communication and teleportation protocols, where they act as continuous-variable carriers of quantum information~\cite{ braunstein1998teleportation}. In photonic quantum computation, PST-based routing supports bosonic encoding schemes with cat states as logical qubits~\cite{ mirrahimi2014dynamically}. Also, preserving cat-state coherence during transport is crucial for quantum error correction and fault-tolerant operations~\cite{leghtas2013stabilizing}. Beyond applications, the PST of cat states provides a controlled framework to study decoherence and the quantum-to-classical transition~\cite{zurek2003decoherence}. Thus, the PST of Schr\"odinger cat states in waveguide networks not only advances integrated quantum photonics but also probes the robustness of macroscopic quantum superpositions.

\subsubsection{TMSV PST}

The evolution governed by Eq.~(\ref{eq:U_general}) is linear and unitary. Hence, the same mechanism applies universally to both DV and CV quantum states. In the DV regime, this manifests as perfect single-photon transfer. In the CV regime, Gaussian states evolve under symplectic transformations that preserve their structure, and the same Fourier modes carry squeezing and correlations.

\begin{figure*}[htbp]
\centering
\includegraphics[width=0.48\linewidth]
{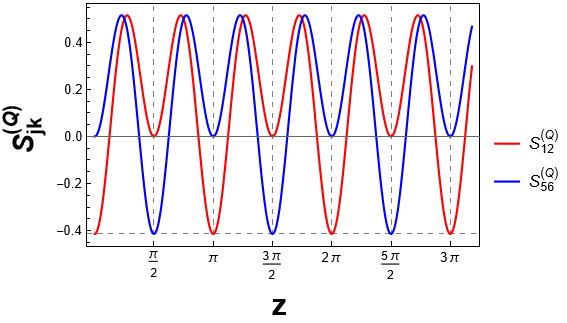}
\put(-208,128){(a)}
\hfill
\includegraphics[width=0.48\linewidth]{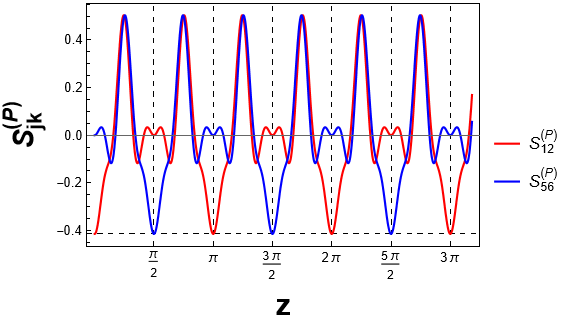}
\put(-202,128){(b)}
\caption{Squeezing factor plots of TMSV for a waveguide quantum network with $N=8$ and $R=3$. Here, we provide TMSV as input to the combination of modes $(1,2)$. (a) The perfect transfer of Q-quadrature squeezing, exactly to the diametrically opposite modes of the network, i.e., $(1,2)\rightarrow(5,6)$. (b) The perfect transfer of P-quadrature squeezing to the antipodal modes of the network, i.e., $(1,2)\rightarrow(5,6)$. We choose the squeezing parameter as $w=0.881374$ and $\theta=0$.}
\label{fig:PST_SQZDall}
\end{figure*}

To demonstrate this, we consider a TMSV state injected into two adjacent modes of the waveguide quantum network with the squeezing operator and the quadrature formalism explained in Appendix B ( see \ref{TMSVFormalism}).  Numerical results for TMSV are given in Fig. (\ref{fig:PST_SQZDall}), where the input modes are $(1,2)$ in an $N=8$ waveguide array. For the coupling range $R=3$, the initial squeezing of combined Q and P-quadrature of $(1,2)$-modes is perfectly transferred to the combined quadratures of $(5,6)$-modes at $Z_{{PST}}$. Both quadratures refocus in accordance with Heisenberg's uncertainty principle, indicating complete reconstruction of the Gaussian state. This establishes that the spectral-collapse condition is not restricted to single-particle dynamics but governs full multimode quantum correlations, including TMSV transport. Our linear combination of quadratures is such that it represents EPR variables, namely, relative position and total momentum \cite{RevModPhys.84.621}. In both EPR variables, perfect squeezing transfer is evident, validating our analytical study.

\begin{figure*}
\centering
\begin{subfigure}
    \centering
    \includegraphics[width=2.2in, height=2.2in]{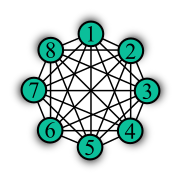}
    \put(-148,135){\textbf{(a)}}
\end{subfigure}
\hfill
\begin{subfigure}
    \centering
    \includegraphics[width=2.2in, height=2.2in]{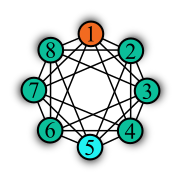}
    \put(-148,135){\textbf{(b)}}
\end{subfigure}
\hfill
\begin{subfigure}
    \centering
    \includegraphics[width=2.2in, height=2.2in]{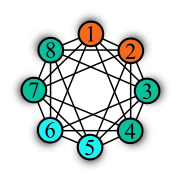}
   \put(-148,135){\textbf{(c)}}
\end{subfigure}

\vspace{0.5cm}

\begin{subfigure}
    \centering
    \includegraphics[width=2.2in, height=2.2in]{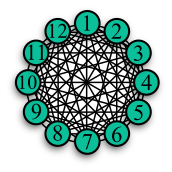}
    \put(-148,135){\textbf{(d)}}
\end{subfigure}
\hfill
\begin{subfigure}
    \centering
    \includegraphics[width=2.2in, height=2.2in]{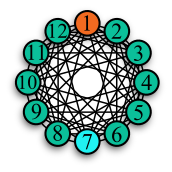}
    \put(-148,135){\textbf{(e)}}
\end{subfigure}
\hfill
\begin{subfigure}
    \centering
    \includegraphics[width=2.2in, height=2.2in]{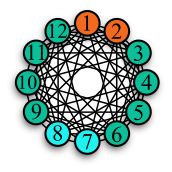}
    \put(-148,135){\textbf{(f)}}
\end{subfigure}
\caption{Sketch of waveguide quantum networks for $N = 8$ and $N=12$ with different interaction ranges and PST combinations. The input modes are shown in orange, and the modes in which the states are perfectly transferred are shown in blue. When input is applied to a network mode at any site $j$, PST occurs at its antipodal mode due to rotational symmetry. For $N=8$ and $N=12$, we depicted the all-to-all coupled model in (a) and (d), the PST of the single photon state, even cat state, and odd cat state in (b) and (e), and the PST of the TMSV state in (c) and (f), respectively at the propagation distance $Z_{PST}$.}
\label{fig:ResultSummaryPictorial}
\end{figure*}

A pictorial representation for PST in $N = 8$, $12$ coupled waveguides is shown in Fig. ~\ref{fig:ResultSummaryPictorial}. This picture highlights the symmetry of PST under the initial conditions of DV and CV, where the excitation injected at any site is transferred to its antipodal partner.

\subsection{Evanescent coupling and robustness of PST}

We now examine the robustness of PST under evanescent coupling conditions. In integrated photonic platforms, waveguide coupling arises from evanescent overlap and decays exponentially with distance. This can be modeled as \cite{mai2014discrete,keil2011classical,tapia2025waveguide}
\begin{equation}
C_r = \mu^r, \quad 0<\mu<1
\end{equation}

\begin{figure*}[htbp]
\centering
\includegraphics[width=0.48\linewidth]
{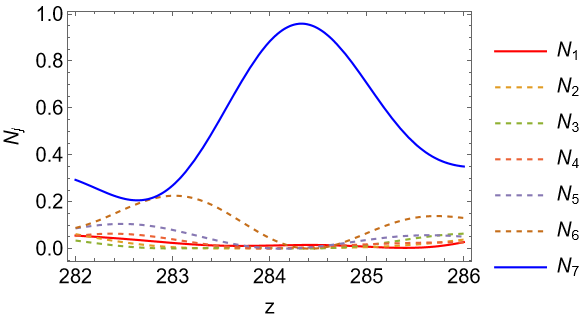}
\put(-213,120){(a)}
\hfill
\includegraphics[width=0.48\linewidth]{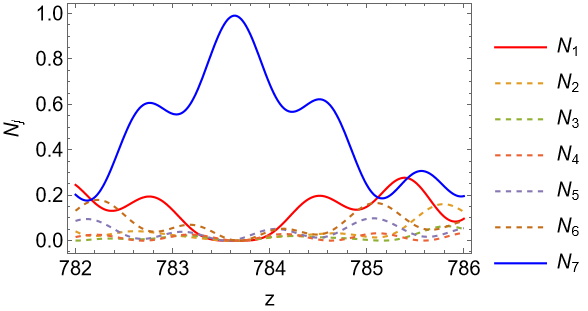}
\put(-213,120){(b)}
\caption{Quantum State transfer of the single photon state under evanescent couplings. Exact PST is lost due to spectral broadening, but near-perfect transfer is achievable at significantly larger propagation distances. (a) $\approx96\%$ transfer at large distance for $N=12$,$ R=6$ and $\mu\approx0.524$, (b) $\approx99\%$ transfer at very large $z$ for $N=12$, $R=5$ and $\mu\approx0.815$.}
\label{fig:PST_EvanESC_all}
\end{figure*}

The parameter $\mu$ is the coupling strength and can be related to physical separation $d$ between the neighbors through the exponential form $\mu=e^{-\kappa d}$, where $\kappa$ depends on waveguide and material properties. Our model yields $r=|j-k|$, where $r$ is the index separation between waveguide modes $j$ and $k$, and $ r$ takes values $ 1,2,...,N/2$ keeping the periodicity. This model assumes uniform spacing and neglects geometric effects, thereby providing an approximate, simplified description of evanescent coupling in circular networks.  Introduction of distance-dependent coupling breaks the uniform coupling condition required for spectral collapse, thereby lifting the three-block eigenvalue structure. The numerical results for state transfers are shown in Fig.~\ref{fig:PST_EvanESC_all}. Although high-fidelity transfer ($>96\%$--$99\%$) can still be achieved for specific $\mu$ values, two key deviations emerge: (1) the transfer is no longer exact due to incomplete phase synchronization, and (2) the required propagation distance increases significantly by many orders of magnitude.

Physically, exponential decay broadens the eigenvalue distribution, introducing phase dispersion across the Fourier modes, as shown in Fig. \ref{fig:ResultSpectra} of Appendix C. As a result, the constructive interference required for PST is only approximate and occurs at much longer propagation lengths.  As larger propagation distances are more prone to loss, they are not favorable for fabrication. Thus, while near-perfect transfer remains achievable, the ideal PST mechanism is intrinsically tied to the uniform long-range coupling condition and is not strictly preserved when coupling strength decays evanescently.

\section{Experimental Feasibility and Implementation}
\label{feasibility}

The realization of uniform long-range coupling up to a finite interaction range is a key experimental requirement for observing PST in the proposed quantum photonic network. While a single auxiliary waveguide can generate global all-to-all coupling, the PST condition derived in this work requires uniform coupling only up to the range $R = N/2 - 1$, with vanishing opposite-site interaction. Such a coupling profile can, in principle, be engineered using auxiliary mediating systems  
\cite{viedma2022auxiliary,  carolan2015universal, wang2020integrated, zhou2023realizing, wu2024modular, sauerwein2023engineering, vijayan2024cavity}. 

We consider an $N$-waveguide quantum photonic network described by annihilation operators $\hat{a}_j$, coupled to $2M$ auxiliary modes $\hat{b}_k$ arranged in symmetric pairs. The total Hamiltonian is
\begin{equation}
\begin{aligned}
H &= \omega \sum_{j=1}^{N} \hat{a}_j^\dagger \hat{a}_j
+ \sum_{k \in \{\pm 1, \dots, \pm M\}} 
\omega_k \hat{b}_k^\dagger \hat{b}_k \\
&\quad \qquad + \sum_{k \in \{\pm 1, \dots, \pm M\}} 
\sum_{j=1}^{N}
\left(
g_k e^{i\frac{2\pi k j}{N}} 
\hat{a}_j^\dagger \hat{b}_k
+ \mathrm{h.c.}
\right)
\end{aligned}
\label{eq:full_H}
\end{equation}

Each auxiliary mode couples uniformly to all waveguides in the network with a controlled phase factor $e^{i\frac{2\pi k j}{N}}$, which can be implemented using propagation-phase engineering, curved auxiliary waveguides, slab-mediated interactions, or synthetic-dimension photonic platforms \cite{yang2025programmable,gehl2017active,ismail2011improved,hilden2025coupled,shan2025non, yoshimi2024efficient}.
\\
In the large detuning regime,
\begin{equation}
\Delta_k = \omega - \omega_k,
\qquad
|\Delta_k| \gg g_k
\end{equation}
The auxiliary modes can be adiabatically eliminated, yielding an effective Hamiltonian
\begin{equation}
H_{\mathrm{eff}} =
\sum_{r} J_r \sum_{j=1}^{N} \hat{a}_{j+r}^\dagger \hat{a}_j
\end{equation}
where $r = i - j$ denotes the separation between sites (defined modulo $N$). The effective coupling profile is given by
\begin{equation}
J_{r} =
\sum_{k \in \{\pm 1, \dots, \pm M\}}
\frac{|g_k|^2}{\Delta_k}
e^{i\frac{2\pi k r}{N}}
\label{eq:J_r_full}
\end{equation}
up to an overall sign depending on the detuning convention. Hermiticity requires $J_{-r} = J_r^*$, which can be enforced by imposing symmetric pairing conditions
\begin{equation}
g_{-k} = g_k^*, \qquad \Delta_{-k} = \Delta_k, \quad k = 1, \dots, M.
\end{equation}
Under these conditions, Eq.~\eqref{eq:J_r_full} reduces to the real-valued form
\begin{equation}
J_{r} =
\sum_{k=1}^{M}
\frac{2|g_k|^2}{\Delta_k}
\cos\left(\frac{2\pi k r}{N}\right)
\label{eq:J_r_real}
\end{equation}
which represents a discrete Fourier synthesis of the coupling profile. By appropriately choosing the auxiliary detunings and coupling strengths, the interaction can, in principle, be engineered (or approximated) to satisfy the PST condition
\begin{equation}
J_r = C \quad (r = 1, 2, \dots, N/2 - 1),
\qquad
J_{N/2} = 0
\label{eq:PST_condition}
\end{equation}
while $J_0$ corresponds to an overall on-site energy shift and is not constrained by the PST requirement.

Defining the coefficients
\begin{equation}
A_k = \frac{2|g_k|^2}{\Delta_k}
\end{equation}
the PST condition reduces to the linear system
\begin{align}
\sum_{k=1}^{M} A_k \cos\left(\frac{2\pi k r}{N}\right) &= C,
\quad r = 1, 2, \dots, N/2 - 1, \\
\sum_{k=1}^{M} A_k (-1)^k &= 0.
\end{align}
This constitutes $N/2$ independent constraints, which admit solutions provided that the number of auxiliary mode pairs satisfies $M \ge N/2$ (i.e., $2M \ge N$ total modes).

The condition $J_{N/2} = 0$ enforces cancellation of the opposite-site coupling and can be satisfied through an appropriate choice of alternating weights $A_k$. In practice, exact realization of the ideal coupling profile may be limited by fabrication and control constraints; however, approximate solutions can be obtained via optimization over experimentally accessible parameters.

Therefore, the required coupling profile for PST can be implemented using a finite number of auxiliary mediating modes, with $2M \sim N$. Similar auxiliary-waveguide and slab-mediated interaction schemes have been proposed for selective mode pumping, long-range coupling, and engineered transport in photonic lattices \cite{viedma2023mode,liu2014method}. Multi-bus waveguide architectures, slab-mediated interactions, and synthetic-dimension photonic systems offer flexible control over coupling profiles, making the proposed PST mechanism experimentally accessible within current integrated photonic technologies \cite{karkar2016network,maier2005experimental}.

\section{Discussion and Conclusion}
\label{discussionAndCon}
We investigated PST in a quantum photonic network with long-range interactions. Owing to the intrinsic properties of the Fourier modes of the proposed network, the eigenvalue spectrum leads to perfect state transport. In particular, we identified that, in networks of arbitrary system size $N=4n$ with uniform long-range couplings and interaction range $R = N/2 - 1$, PST is enabled by a highly structured spectral collapse into three distinct eigenvalue blocks, the important one being the $N/2$-manifold of zero Fourier modes. These zero Fourier modes ensure the protected propagation of the input quantum states throughout the propagation distance. The spectral degeneracy ensures the required phase conditions across all Fourier modes, leading to complete constructive interference at the target site and destructive interference at rest sites.

Our analytical and numerical results for a single photon state demonstrate that exact PST occurs only under uniform long-range couplings with an interaction range $R = N/2 - 1$ in systems satisfying $N=4n$, whereas PST occurs at the diametrically opposite mode. Deviations from this condition distort the Fourier mode structure, leading to rapid degradation of transfer fidelity. This highlights the role of spectral structure in achieving PST rather than local coupling strength or extended coupling ranges. Beyond a specific coupling range, even with higher-order interactions, PST is not observed. Another notable feature of our results is the PST period. We observe PST to antipodal mode at shorter propagation lengths, even for larger systems, with the shortest PST distance of $ \pi/2$, and a revival in input mode at distance $ \pi$ (both for unit coupling strength).

Moreover, the same mechanism extends beyond the single-photon regime. Due to the linear and unitary nature of the evolution, both DV and CV states undergo identical mode transformations. We further investigated Schr\"odinger cat states and found that even and odd cat states exhibit PST at the usual PST period, whereas YS cat states do not exhibit unit fidelity. Schr\"odinger cat state results suggest that the PST mechanism is sensitive to the phase of the input cat states. We report that optimal conditions for achieving higher fidelity in the cat state transfer correspond to a small average photon number and an even cat state. Extending our analysis to two-mode squeezed-vacuum inputs, we demonstrated perfect transport of quadrature squeezing and Gaussian correlations, confirming that the spectral-collapse condition governs the routing of nonclassical resources in multimode systems.

We further examined the impact of evanescent coupling, modeled by exponentially decaying interaction strengths, for a single-photon state. In this case, the Fourier mode spectrum is no longer preserved, resulting in a broadened eigenvalue distribution. Although near-perfect and high fidelity transfer remains achievable, it occurs only at significantly larger propagation distances, which are experimentally unfavorable due to losses and device constraints. This highlights a fundamental challenge in achieving PST in experimentally easy-to-realize, evanescently coupled, larger waveguide systems.  As discussed in the feasibility section, these constraints can be addressed using auxiliary systems to engineer ideal transport conditions.

Overall, our results establish waveguide-based quantum photonic networks as a versatile platform for studying quantum transport. The spectral perspective developed here is broadly applicable to other systems, including photonic lattices, spin chains, and engineered quantum graphs \cite{mulken2011quantumgraphs, childsg2009universal, rodriguezlara2022su2}. The antipodal transfer property of our system, as shown in its pictorial representation, can be leveraged for applications such as quantum switching \cite{yung2011spin}. By utilizing PST to move quantum states into a target mode, our system enables quantum routing that is significantly faster and more efficient than traditional architectures. These findings open avenues for exploring disorder and defect-induced effects, controlled coupling engineering in circular networks, and scalable routing of multimode quantum correlations in integrated quantum architectures.

\section*{Acknowledgements}

S.M. would like to thank CSIR India (09/0263(15337)/2022-EMR-1) for the fellowship during this work. P. A. B. thanks the financial support of CNPq (Conselho Nacional de Desenvolvimento Cient\'ifico e Tecnol\'ogico). A.R. gratefully acknowledges support from the Anusandhan National Research Foundation (ANRF) of the Department of Science and Technology (DST), India, through the Advanced Research Grant (ARG) (File Number: ANRF/ARG/2025/009823/PS).

\subsection*{Appendix A: Spectral collapse in circular arrays with uniform long-range coupling}

Here, we present the analytical steps leading to the spectral collapse used in Sec.\,\ref{results}. We consider an $N$-waveguide quantum network with uniform long-range coupling $C_r=C$ and interaction range $R=N/2-1$ under periodic boundary conditions. In the single-excitation subspace, the Hamiltonian becomes
\begin{equation}
H = \sum_{r=1}^{N/2-1} C \sum_{j=0}^{N-1}
\left(\,
|j\rangle \langle j+r| + |j+r\rangle \langle j|
\,\right)
\end{equation}
Due to translational symmetry, the eigenstates are discrete Fourier modes
\begin{equation}
|\lambda_p\rangle = \mathcal{S}_{j,p}\,|j\rangle
\end{equation}
which diagonalize the Hamiltonian. The corresponding eigenvalues are
\begin{equation}
\lambda_p = 2C \sum_{r=1}^{N/2-1}
\cos\!\left(\frac{2\pi p r}{N}\right)
\label{lambda_appendix}
\end{equation}
To simplify the sum, we use the discrete Fourier identity
\begin{equation}
\sum_{r=0}^{N-1} e^{\frac{2\pi i p r}{N}} = 0,
\qquad p\neq0
\end{equation}
Separating the terms $r=0$ and $r=N/2$ and using cosine symmetry gives
\begin{equation}
1 + (-1)^p + 2\sum_{r=1}^{N/2-1}
\cos\!\left(\frac{2\pi p r}{N}\right) = 0
\end{equation}
which leads directly to Eq.~ \ref{eqn:LambdaPGen}. Additionally, the $p=0$ mode gives $\lambda_0 = C(N-2)$. Thus, the spectrum collapses into three distinct eigenvalue blocks as given in Eqs. (10-12). This highly structured spectrum produces $N/2$ degenerate zero Fourier modes and a single high-energy mode separated from the rest of the spectrum. Such spectral collapse enables phases across the Fourier modes, which are responsible for the condition discussed in Sec.\,\ref{results}.

\subsubsection*{Effect of opposite-site coupling}
If opposite-site coupling ($r=N/2$) is included, the eigenvalues become
\begin{equation}
\lambda_p =
2C\sum_{r=1}^{N/2-1}
\cos\!\left(\frac{2\pi p r}{N}\right)
+ C\cos(\pi p)
\end{equation}
Using the previous identity, this simplifies to
\begin{equation}
\lambda_0 = C(N-1),
\qquad
\lambda_p = -C \quad (p\neq0)
\end{equation}
which is a completely different spectrum from the earlier case. In this case, the probability amplitude is given by $\langle p|U(Z_{\text{PST}})|q\rangle$
\begin{equation}
\langle p|U(Z_{\text{PST}})|q\rangle
= \frac{1}{N} \Big[ e^{-iC(N-1)z}-e^{iCz}\Big]
\label{eqn:NOTPST}
\end{equation}
where $p\neq q$ is assumed. Eq. (\ref{eqn:NOTPST}) demonstrates that PST cannot occur for $N>2$.

\begin{figure*}
\centering
\begin{subfigure}
    \centering
    \includegraphics[width=3.1in, height=1.9in]{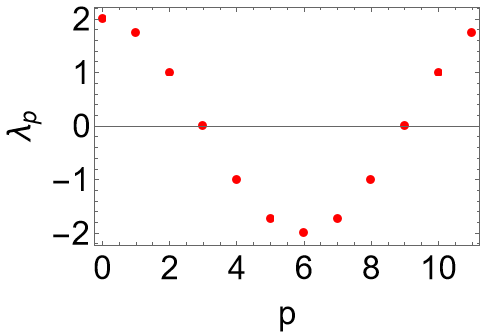}
    \put(-100,122){(a)}
\end{subfigure}
\hfill
\begin{subfigure}
    \centering
    \includegraphics[width=3.5in, height=1.9in]{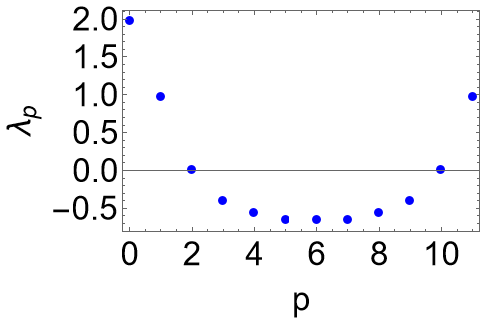}
    \put(-100,122){(b)}
\end{subfigure}
\vspace{0.5cm}
\begin{subfigure}
    \centering
\includegraphics[width=3.1in, height=1.9in]{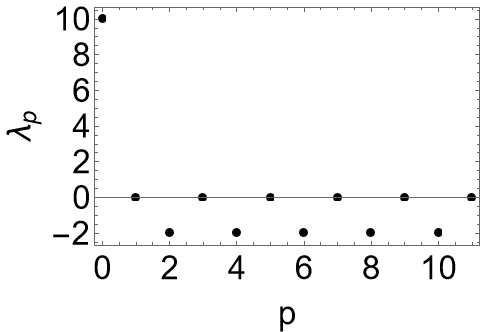}
\put(-100,122){(c)}
\end{subfigure}
\hfill
\begin{subfigure}
    \centering
\includegraphics[width=3.1in, height=1.9in]{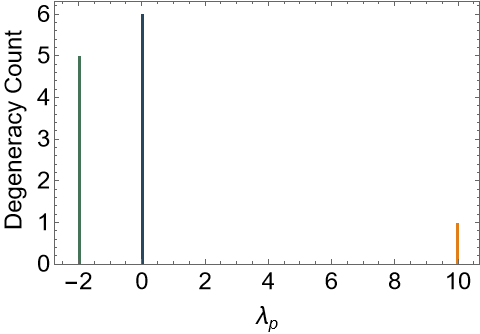}
\put(-100,122){(d)}
\end{subfigure}
\caption{For a waveguide quantum network with $N=12$ and different coupling profiles. (a) For nearest neighbor coupling only, the spectrum is exactly of cosine form and has various energy modes (b) for evanescent coupling with range $R=6$ for $\mu=0.5$, showing a broadened spectrum, (c) highly structured and degenerate supermode spectra for PST case, (d) histogram for PST case showing degeneracy count of energy values.}
\label{fig:ResultSpectra}
\end{figure*}
\begin{figure*}
\centering
\includegraphics[width=0.38\linewidth]
{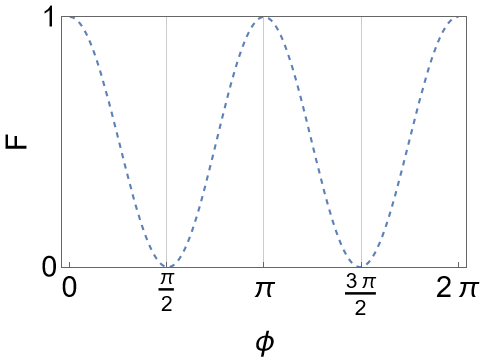}
\put(-45,114){(a)}
\hfill
\includegraphics[width=0.575\linewidth]{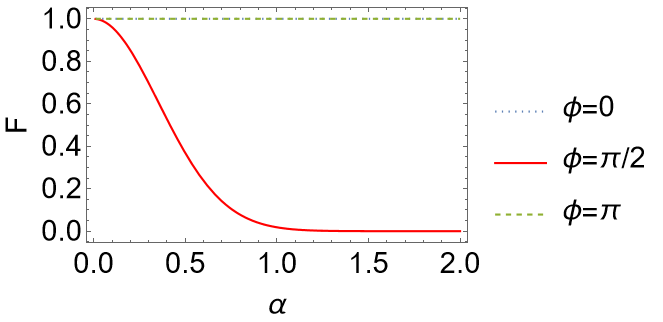}
\put(-143,112){(b)}
\caption{For a twelve-waveguide array at PST condition, we observe fidelity values varying with $\alpha$ and $\phi$ (a) $F$ varying with $\phi$ at $Z_{PST}=\pi/2$ for $\alpha=\sqrt{2}$ (b) $F$ at $Z_{PST}=\pi/2$ for varying $\alpha$ values from $0.01$ to $2$.}
\label{fig:catStateFgeneral}
\end{figure*}

\subsection*{Appendix B: Two-mode squeezed vacuum and quadrature formulation}
\label{TMSVFormalism}
In this appendix, we present the covariance-matrix formalism \cite{weedbrook2012gaussian, adesso2014continuous} and quadrature definitions used to analyze two-mode squeezing and its propagation in the waveguide array. The input state is a two-mode squeezed vacuum (TMSV) injected in selected modes, while the remaining modes are initialized in vacuum. The two-mode squeezing operator acting on modes $m$ and $n$ is
\begin{equation}
\hat{T}_{mn}(\xi)=\exp\!\left[       \xi\, \hat{a}_m^\dagger \hat{a}_n^\dagger-\xi^*\, \hat{a}_m \hat{a}_n  \right]
\end{equation}

where $\xi = w \, e^{i\theta}$ is the squeezing parameter, with $w$ denoting the squeezing strength and $\theta$ the squeezing phase. Acting on the vacuum, it generates the two-mode squeezed vacuum state $|\psi_{\mathrm{TMSV}}\rangle = \hat{T}_{mn}(\xi)|0,0\rangle.$ The quadrature for individual modes is defined as
\begin{equation}
\hat{Q}_{j} = \frac{1}{\sqrt{2}}(\hat{a_j} + \hat{a_j}^\dagger)
\end{equation}
\begin{equation}
\hat{P}_{j} = \frac{1}{i\sqrt{2}}(\hat{a_j} - \hat{a_j}^\dagger)
\end{equation}

To characterize the squeezing in TMSV, the combined quadrature of the superposition of two modes defined in ref.~\cite{RevModPhys.84.621}, generally named as relative position and total momentum and called EPR variables, as $\hat{Q}_{jk}=\frac{1}{\sqrt{2}}(\hat{Q}_{j} - \hat{Q}_{k})$ and $\hat{P}_{jk}=\frac{1}{\sqrt{2}}(\hat{P}_{j} + \hat{P}_{k})$ which in form of individual mode operators will become as-
\begin{equation}
\hat{Q}_{jk} = \frac{1}{2}(\hat{a}_{j} + \hat{a}_{j}^\dagger-\hat{a}_{k} - \hat{a}_{k}^\dagger)
\end{equation}
\begin{equation}
\hat{P}_{jk} = \frac{1}{2i}(\hat{a}_{j} - \hat{a}_{j}^\dagger+\hat{a}_{k} - \hat{a}_{k}^\dagger)
\end{equation}
these combined quadratures satisfy the canonical commutation relation $[\hat{Q}_{jk},\hat{P}_{jk}]=0$.  With this normalization, the variance for the vacuum state is
\begin{equation}
\langle (\Delta \hat{Q}_{jk})^2 \rangle
=
\langle (\Delta \hat{P}_{jk})^2 \rangle
=
\frac{1}{2}
\end{equation}

In the basis of individual modes, with quadratures $\hat{\xi} = (\hat{Q}_{1},\hat{P}_{1},...,\hat{Q}_{N},\hat{P}_{N})$, the solution of the propagation Eq. (\ref{eq:HEOM}) can be written as $\hat{\xi}(z)=\mathcal{M}(z)\hat{\xi}(0)$, with $\mathcal{M}(z)$ a symplectic matrix that encapsulates all the information about the propagation of the quantum state in the network.

Gaussian states can be fully characterized by their first moments and covariance matrices. The initial covariance matrix with input as the TMSV state in two adjacent modes and the vacuum state in the remaining $N-2$ modes of the network can be written as
\begin{equation}
V_i =
\begin{pmatrix}
V_{\mathrm{TMSV}} & 0 \\
0 & V_{I}
\end{pmatrix}
\end{equation}
where $V_{\mathrm{TMSV}}$ is the $4\times4$ covariance matrix of the TMSV state and $V_{I}$ is the covariance matrix of the vacuum modes, in shot noise units, it can be written as $V_{I} = \frac{1}{2}I$, with $I$ as the identity matrix of dimension $(2N-4)\times(2N-4)$. The final covariance matrix for any propagation length $z$ is given by $V_f = \mathcal{M}(z) V_i \mathcal{M}^T(z)$, evolution of variances $V(\xi_{j},\xi_{j})$ and quantum correlations $V(\xi_{j},\xi_{k})$ at any length can be obtained from the elements of this matrix.

The two-mode squeezing between any pair of modes $(j,k)$ is quantified through the squeezing factor defined as
\begin{equation}
S_{jk}^{(Q)} =
\langle (\Delta \hat{Q}_{jk})^2 \rangle - 1/2
\end{equation}
\begin{equation}
S_{jk}^{(P)} =
\langle (\Delta \hat{P}_{jk})^2 \rangle - 1/2
\end{equation}
A negative value of the squeezing factor indicates quadrature squeezing
\begin{equation}
S_{jk}<0
\end{equation}
while $S_{jk}=0$ corresponds to the vacuum level and $S_{jk}>0$ indicates quantum noise. In the present work, the TMSV is injected into the first two modes, while the remaining modes are maintained under vacuum. The symplectic matrix generates the evolved covariance matrix ${V}_f$, from which the squeezing factors $S^{Q/P}_{12}$ and $S^{Q/P}_{56}$ are computed. This allows direct tracking of the transfer and redistribution of two-mode squeezing across the waveguide array as a function of propagation distance. This formulation is fully consistent with Gaussian-state quantum optics and provides a direct connection between covariance matrix elements and measurable quadrature squeezing.

\subsection*{Appendix C: Spectral Analysis For Different Coupling Profiles}
For three cases of coupling profiles, we plot supermode spectra, and it is evident from Fig. (\ref{fig:ResultSpectra}) that the spectrum is highly structured and degenerate in the PST case. 

\subsection*{Appendix D: Schr\"odinger cat state fidelity}
From Eq. (\ref{PSTNegativePhase}), it is clear that at a distance $z = Z_{PST}$, the transition amplitude $U_{jl}(z)$, between antipodal modes, satisfies $U_{jl}(Z_{PST})=-1$. Thus, the expression for the fidelity shown in Eq.~\eqref{FidCat} takes the form 
\begin{equation}
    F_{PST}= \Bigg| 2\,\mathcal{N}^2 e^{-|\alpha|^2} \Big( e^{-|\alpha|^2}+ \, \cos{\phi} \,e^{   \,|\alpha|^2}\Big) \Bigg|^2.
    \label{FCAtStateAppndx}
\end{equation}
For $\phi=0$ or $\pi$, the $\cos{\phi}$ term contributes with $\pm1$, which explains why we get higher fidelity values for even cat states. If $\phi=\pi/2$, the $\cos{\phi}$ term does not contribute and the fidelity becomes
\begin{equation}
    F_{PST}(YS)= |2\,\mathcal{N}^2 e^{-2|\alpha|^2}|^2,
    \label{FCAtStateAppndx}
\end{equation}
which exponentially decreases with $\alpha$, as Fig.~\ref{fig:catStateFgeneral} demonstrates. The reduced values for the fidelity in the case of YS cat states and higher fidelity for even cat states originate from the phase-sensitive superposition terms in the fidelity expressions. 

Notice that the $\alpha = 0$ case corresponds to the vacuum field for a coherent state, but the odd cat state is not defined at $\alpha = 0$ because the antisymmetric superposition gives $\ket{\alpha} - \ket{\alpha} = \ket{0}-\ket{0} = 0$, which is the zero vector (not the vacuum state). If one takes the limit $\alpha \rightarrow 0$, the odd cat state tends to the single-photon Fock state $\ket{1}$, not to the vacuum. This is why the value $\alpha = 0$ is not included in the plot of Fig. \ref{fig:catStateFgeneral}.

\color{black}

\bibliography{apssamp}
\end{document}